%% file: LayertuneSO_arXiv_submit.tex
\documentclass[aps,prl,twocolumn,groupedaddress]{revtex4-1}

\usepackage{multirow}
\usepackage{mathrsfs,amsmath} 
\usepackage{amsfonts}
\usepackage{amssymb}
\usepackage{graphicx}
\usepackage{epstopdf}
\usepackage{setspace}
\usepackage[toc,page]{appendix} 
\usepackage{placeins}
\usepackage{chngcntr}
\usepackage{titlesec}
\usepackage{hyperref}

\begin{document}

\author{Jun Yong Khoo${}^1$, Alberto F. Morpurgo${}^2$, Leonid Levitov${}^1$}
\affiliation{${}^1$ Department of Physics, Massachusetts Institute of Technology, 77 Massachusetts Avenue, Cambridge, Massachusetts 02139, USA\\
${}^2$ Department of Quantum Matter Physics (DQMP) and Group of Applied Physics (GAP), University of Geneva, 24 Quai Ernest-Ansermet, CH1211 Geneve 4,
Switzerland}

\title{On-Demand Spin-Orbit Interaction from Which-Layer Tunability in Bilayer Graphene}

\keywords{Bilayer graphene, spin-orbit interaction, gate-tunability, intrinsic valley-Hall conductivity, topological phase transitions}

\begin{abstract}
Spin-orbit interaction (SOI) that is gate-tunable over a broad range is essential to exploiting novel spin phenomena. Achieving this regime has remained elusive because of the weakness of the underlying relativistic coupling and lack of its tunability in solids. Here we outline a general strategy that enables exceptionally high tunability of SOI through creating a which-layer spin-orbit field inhomogeneity in graphene multilayers. An external transverse electric field is applied to shift carriers between the layers with strong and weak SOI. Because graphene layers are separated by sub-nm scales, exceptionally high tunability of SOI can be achieved through a minute carrier displacement. A detailed analysis of the experimentally relevant case of bilayer graphene on a semiconducting transition metal dichalchogenide substrate is presented. 
In this system, a complete tunability of SOI amounting to its ON/OFF switching can be achieved. New opportunities for spin control are exemplified with electrically driven spin resonance and topological phases with different quantized intrinsic valley Hall conductivities.
\end{abstract}

\maketitle

Spin-orbit interaction (SOI), tunable on demand over a wide range of values can provide access to a wide variety of interesting spin transport phenomena. One popular strategy of achieving tunable SOI relies on directly tuning the SOI using an applied electric field. This approach proved successful in various instances such as tuning Rashba-type SOI in two-dimensional semiconducting systems\cite{GaAs,InGaAs,Oxide1,Oxide2} and Ising-type SOI in transition metal dichalcogenides (TMDs)\cite{SV1}. However, in all these cases the range of values in which SOI could be tuned has been relatively small because of the relativistic nature of SOI.

We propose graphene multilayers as a vehicle to achieve an on-demand SOI that is free from these limitations. The first step involves engineering an environment with a spatially inhomogeneous spin-orbit field\cite{app1,app2}, which is e.g. high on one layer and low on the adjacent layer. In such a system, through applying transverse electric field, carriers can be shifted between layers with strong and weak SOI. This renders the SOI strength felt by these carriers strongly dependent on the which-layer charge polarization. Some aspects of this scheme resemble gate-tunable Zeeman coupling demonstrated in Ref.~\citenum{gfactor}. The advantage of such an indirect approach to tuning SOI is that it disassociates the applied electric field from the spin-orbit field. 
The atomic scale separation between graphene layers then ensures an exceptionally high tunability that is achieved through a minute carrier displacement.

\begin{figure}
\hspace{-1em}\includegraphics[scale=0.63]{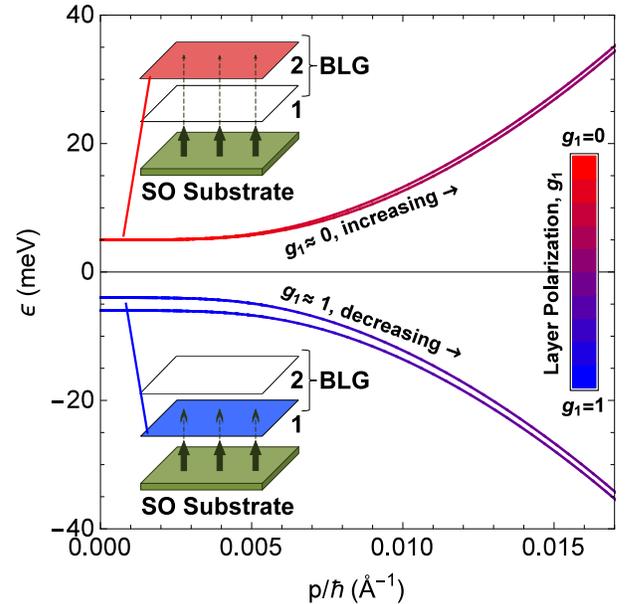}
\caption{\label{schematic} Low-energy band structure of a biased bilayer graphene (near $K_+$ point) with interlayer bias $U = 10$ meV including induced SOI $H_{\text{SO}}^{\text{eff}} =  \frac{1}{2} g_1^{(\zeta)} \lambda s_z$ with $\lambda = 2$ meV. The color of the lines indicates the layer occupancy $g_{1,2}^{(\zeta)}$ of their corresponding eigenstates given by eq \ref{Eqn.1.3} in the main text. A similar situation occurs near the $K_-$ point. Different sizes of arrows in the insets depict the difference in strength of the induced SOI experienced by carriers in layer 1 and 2 as a result of their proximity from the substrate.}
\end{figure}

We illustrate this idea in the specific context of bilayer graphene (BLG) on a TMD substrate such as WS$_2$. Implementing a strongly tunable SOI in such graphene-based systems is highly desirable due to the high mobility of carriers in graphene that is preserved by these atomically flat and chemically inert substrates\cite{mobility}. In this configuration, the spatially inhomogeneous spin-orbit field simplifies to an ON/OFF which-layer field -- only the layer adjacent to the TMD acquires from it an interfacially-induced Rashba SOI and Ising SOI. Our proposal builds on previous work, which established that strong interfacial SOI in the meV range can be induced in individual graphene layers\cite{MLGWS2,BLGTMD,MLGWS2SG,MLGWS2Shi,MLGTMD}. As we will show, the low-energy carriers experience an effective SOI that has an enhanced gate-tunability to the extent of \textit{complete gate-tunability}, i.e. it can be switched on and off by applying a transverse electric field of moderate strength ($\sim$ mV/\AA).

Furthermore, the robust high-frequency response of graphene extending up to $\sim$ 100 GHz\cite{dynamic1} can enable a range of novel time-dependent spin phenomena. Indeed, because applying a transverse field in BLG directly alters the wavefunctions of its carriers, gate-tunable SOI possesses full quantum coherence. Quantum-coherent tunability enables coherent manipulation of carrier spin degrees of freedom, becoming particularly interesting if the SOI Hamiltonian can be modulated on the carrier transport time scales. 
As an illustration of this new capability, we discuss the electric-dipole spin resonance (EDSR) that can be driven though an application of a transverse AC electric field\cite{EDSR}. Quantum-coherent tunability also enables direct control of the electron Bloch Hamiltonian and Bloch bands, giving access to gate-tunable Berry phase and band topology of Bloch electrons. We illustrate these new opportunities by considering BLG sandwiched between TMD layers, a system that provides gate-switching between
topologically distinct phases with different values of the intrinsic valley Hall conductivity.

The essential aspects of the which-layer approach can be illustrated by a model of a Bernal-stacked BLG in a transverse electric field, which for the sake of simplicity only accounts for the low-energy subspace of electronic states. Microscopically, the interlayer bias potential $U$ introduces an asymmetry between the A sublattice of layer 1 and B sublattice of layer 2, denoted below as A1 and B2. Crucially, the interlayer bias breaks the layer-occupation symmetry\cite{BLCcap}. This behavior is captured by the two-band (spin-degenerate) Hamiltonian describing the low-energy subspace,
\begin{equation}\label{eq:BLG_low-energy}
H^{2 \times 2}_{\text{BLG}} = \left( \begin{array}{cc}
-\frac12U & -\frac{1}{2m} \pi ^{\dagger 2} \\
-\frac{1}{2m} \pi ^2 & \frac12 U
\end{array} \right) , \quad  \pi = \tau _z p_x + i p_y.
\end{equation}
Here $\vec{p}$ is the momentum measured relative to the $K$ and $K'$ points of the Brillouin zone, which we will henceforth refer to as $K_+$ and $K_-$ ($\tau _z = \pm1 $) respectively, and $-\frac12U$ and $\frac12U$ are the potentials on layers $1$ and $2$. The spectrum of $H^{2 \times 2}_{\text{BLG}}$ is given by 
\begin{equation}\label{eqn1.2}
\varepsilon _{\zeta} (p) = \frac{ \zeta}{2} \sqrt{U^2 + \frac{p^4}{m^2}}.
\end{equation}
Here $\zeta = \pm1$ refers to the conduction or valence band respectively, and the interlayer bias $U$ incorporates the capacitance corrections\cite{BLCcap}. When $U \neq 0$, the wavefunctions of electronic states are asymmetric in the layer occupancy:
\begin{equation}\label{Eqn.1.3}
g_{1,2}^{(\zeta)} = |\psi_{\zeta} ^{A1,B2} (p)| ^2 = \frac{1}{2} \mp \frac{U}{4\varepsilon _{\zeta}(p)},
\end{equation}
where the minus and plus signs correspond to the layers 1 and 2, respectively.
The extent of layer polarization for each of these states is therefore directly controlled by $U$. The carriers with specific layer polarization can be accessed in an energy-resolved manner through doping\cite{BLGtune}.

Next, we discuss how a layer-dependent SOI is engineered using a proximal TMD layer, e.g. a TMD multilayer with strong SOI such as WS$_2$ which serves as the substrate for the BLG. We expect carriers in layer 1 (blue) that is adjacent to the TMD substrate to acquire an interfacially-induced SOI (see Fig.\ref{schematic}). Carriers in non-adjacent layer 2 (red) are coupled to substrate only indirectly, through interlayer hopping. This phenomenology of interfacially-induced SOI is supported by recent studies\cite{MLGWS2, MLGTMD} on monolayer graphene (MLG) on TMD substrates with strong SOI. A simple model of MLG experiencing an enhanced SOI due to the TMD substrate can be described by a low-energy Hamiltonian
\begin{equation}\label{Eqn.1.4}
\delta H_{\text{MLG}} = \frac{\Delta}{2}\sigma _z + \delta H_\text{Ising} + \delta H_\text{R},
\end{equation}
treated as a perturbation to the MLG Hamiltonian near the $K_\pm$ points. Here $\delta H_\text{Ising} = \frac{\lambda}{2}\tau _z s_z$ and $\delta H_\text{R} = \frac{\lambda _\text{R}}{2}(\tau _z \sigma _x s_y - \sigma _y s_x )$, where we use $\sigma _i$ and $s_i$ to denote the Pauli matrices corresponding to the A and B sublattices, and to spin degrees of freedom, respectively. The term $\delta H_\text{Ising}$ has the form of Ising SOI and originates from the Ising SOI inherently present in the TMD substrate. The term $\delta H_\text{R}$ has the form of Rashba SOI in graphene. The term $\frac{\Delta}{2}\sigma _z$ originates from sublattice asymmetry; it is comparatively small in practice and can be ignored in most cases.

The consequence of introducing layer-dependent SOI in BLG can be illustrated by considering a simple model in which the interfacial SOI described by $\delta H_{\text{MLG}}$ only affects the carriers localized in layer 1 and is negligible for carriers localized in layer 2. To see how this modification allows for a switchable SOI, consider the limit of weak Ising SOI, $|U| \gg |\lambda|$, and with $\Delta = \lambda _\text{R} = 0$. To first order in perturbation theory, we neglect the influence of $\delta H_{\text{MLG}}$ on the electronic states and find the spin-split low-energy bands 
\begin{equation}\label{Eqn.1.5}
\delta \varepsilon_{\zeta, s=\uparrow, \downarrow} = \pm \frac{\lambda}{2} \tau_z g_1^{(\zeta)} ,
\end{equation}
where the energy shifts of sign plus and minus correspond to the $s = \uparrow$ and $s = \downarrow$ states respectively. The spin splitting in eq \ref{Eqn.1.5} is of opposite sign for different valleys, $\tau_z = \pm1$, as required by time reversal symmetry.

\begin{figure}
\includegraphics[scale=0.36]{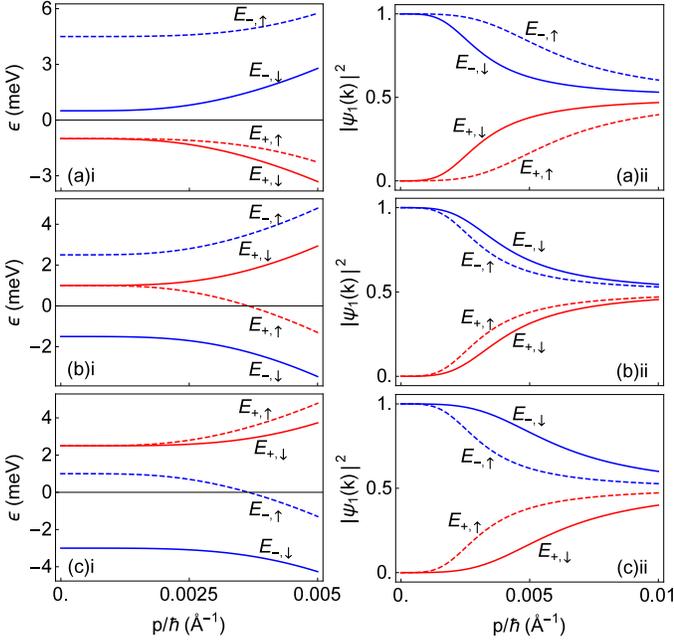}
\caption{\label{figband}
Four lowest energy bands (left panel) and corresponding layer polarization (right panel) obtained from eq \ref{Eqn.2.2}. Upon increasing $U$, the system first transitions from an insulating phase (a) to a semi-metallic phase (b) and then to a different insulating phase (c) that has the order of the bands inverted compared to (a); labels correspond to the phase labels in Fig.\ref{phase}. Values used are (a) $U = -2\,{\rm meV}$, (b) $U = 2\,{\rm meV}$, (c) $U = 5\,{\rm meV}$ with other parameters fixed as $\Delta = 3\,{\rm meV}$, $\lambda = 4\,{\rm meV}$, and $\lambda _\text{R} = 2\,{\rm meV}$.}
\end{figure}

Bands with different spin splitting can be accessed in a dual-gated system in which there is independent control over interlayer bias and doping: the induced SOI is turned on by hole doping and turned off by electron doping. 
Indeed, at small dopings, since the Fermi momentum is small, there is a correlation between which band a carrier is from and which layer of the BLG the carrier predominantly occupies. For positive interlayer bias $U > 0$, carriers from the conduction band are fully localized on layer 2, $g_1 ^{(+1)} \approx 0$, while carriers from the valence band are fully localized on layer 1, $g_1^{(-1)} \approx 1$. It follows from eq \ref{Eqn.1.5} that in this case the Ising SOI is only present for holes ($\delta \varepsilon_{-1,s=\uparrow, \downarrow} \approx \pm \frac{1}{2} \lambda \tau_z$) and is absent for electrons ($\delta \varepsilon_{+1,s} \approx 0$), as illustrated in Fig.\ref{schematic}; the situation is reversed when $U < 0$ so that the Ising SOI is only present for electrons and is absent for holes (see Fig.\ref{figband}(a) and (c)). The contrast between the ON and OFF states fades away quickly as doping increases, since at large momenta the electron wavefunctions are split nearly equally between both layers (see Fig.\ref{figband} right panel).

One interesting consequence of which-layer tunability is that the spin splitting (eq \ref{Eqn.1.5}) acquires a dependence on the interlayer bias $U$. Crucially, the states in eq \ref{Eqn.1.5}, while having opposite spin projections, have identical orbital structure. It is therefore possible to view the spin splitting in eq \ref{Eqn.1.5} as being due to an effective magnetic field applied transverse to the BLG plane. Because of the dependence of the layer occupancy on the interlayer bias $U$ (eq \ref{Eqn.1.3}), this effective $B$ field is gate-tunable and therefore defines a new form of spin-electric coupling.

As an illustration of the new capabilities endowed on the system by such spin-electric coupling, we discuss spin resonance of an EDSR type driven by a time-dependent gate voltage $U(t)$. We consider an external static magnetic field of strength such that the Zeeman energy exceeds the interfacially induced spin splitting, eq \ref{Eqn.1.3}, for the sake of simplicity taking the field to be applied parallel to the BLG plane. The carrier spin dynamics is then governed by an effective Hamiltonian
\begin{equation}
H_{\text{EDSR}} = \frac{1}{2} \epsilon_{\rm Z} s_x + \frac{1}{2}\tilde{\lambda} (t) \tau_z s_z,
\end{equation}
where the time-dependence $\tilde{\lambda} (t)=\lambda g_1^{(\zeta)}(t)$ originates from gate-tunable Ising SOI. Here $\epsilon_{\rm Z} = g \mu_{\rm B} B $ is the Zeeman energy, $\mu_{\rm B}$ is Bohr's magneton and, without loss of generality we consider the static magnetic field applied along the $x$ direction, $\vec{B} = B \hat{x}$.

To achieve EDSR, the interlayer bias $U(t)$ should not at any point in time close the gap between the valence and conduction bands so that the carrier orbital wavefunctions remain unchanged. Without loss of generality, we consider the case for which $\lambda > 0$ and $U(t) = U_0 + U_1 \cos \omega t$ with $U_0 > \lambda$ and $U_1 \ll U_0$. The resulting time-dependent Ising SOI experienced by the conduction band carriers with momentum $p$ varies with time as $\tilde{\lambda} (t) \simeq \frac{\lambda p^4}{4 m^2 U^2(t)}$. This time-dependent Ising SOI will thus act as an oscillating field which induces transitions between the Zeeman  states $s = |\leftarrow \rangle$ and $s = |\rightarrow \rangle$. Consequently, EDSR is achieved by matching the frequency of the time-dependent Ising SOI to the Zeeman energy, $\hbar \omega = \epsilon_{\rm Z}$. Note that while the Ising SOI has opposite signs at the $K_{\pm}$ valleys, both SOI couplings cause the spin projection on the $x$-axis to evolve with the same time dependence. In the absence of intervalley coupling, the effects of EDSR originating from both valleys add up constructively, resulting in doubling of the spin polarization signal.

While the simple analysis of interfacially induced SOI presented above qualitatively captures the essential physics, it is instructive to develop a more precise and complete description of the system near the $K_\pm$ points. That can be done by directly adding $\delta H_{\text{MLG}}$, eq \ref{Eqn.1.4}, to the layer-1 subspace of the BLG tight-binding Hamiltonian,
\begin{equation}\label{Eqn.2.2}
H_{\text{eff}} = H_{\text{BLG}} \otimes \mathtt{1}^{(s)} + \mathcal{P}_1 \delta H_{\text{MLG}}\mathcal{P}_1.
\end{equation}
Here $\mathcal{P}_{i=1,2}$ is the operator that projects onto the layer-$i$ subspace and $\mathtt{1}^{(s)}$ is the $2 \times 2$ identity matrix of the spin degrees of freedom. The Hamiltonian $H_{\text{BLG}}$ describes Bernal-stacked BLG, in which two MLG layers are stacked such that the B sublattice in one layer (B1) is vertically aligned with the A sublattice on the other (A2). As is well known, the strongest interlayer coupling $\gamma_1$ in this stacking configuration occurs between the B1 and A2 sites. As a result the low-energy states near the $K_\pm$ points are strongly localized over the A1 and B2 sublattices, as described by the low-energy Hamiltonian considered above, eq \ref{eq:BLG_low-energy}. The Hamiltonian $H_{\text{eff}}$ can be numerically solved to obtain the band structure. The four low-energy bands obtained in this way are shown in Fig.\ref{figband} for three different values of interlayer bias $U$ corresponding to three different phases discussed below.

To gain insight in the different regimes accessible though varying $U$, we derive the low-energy Hamiltonian in the A1/B2 subspace perturbatively in $\frac{1}{\gamma _1}$(see Supporting Information). Since $\left[ H_{\text{eff}}, s_z \right]=0$ at $p = 0$, the quantity $s_z$ is a good quantum number and can be used to label states and associated bands. To order $\frac{1}{\gamma _1^2}$, the energy levels at $p = 0$ are
\begin{align}\nonumber 
\left. E_{\zeta = +1,s}\right|_{p=0} &\simeq  \frac{U}{2} \hspace{1em}\text{(doubly degenerate, } s = |\uparrow \rangle, |\downarrow \rangle), \label{Eqn.peq0b} \\
\left. E_{\zeta = -1,s}\right|_{p=0} &\simeq  - \frac{U}{2} + \frac{\Delta}{2}+ \delta \lambda^* \pm \left(\frac{\lambda}{2} - \delta \lambda^*\right)\tau_z, \quad 
\\ \nonumber
\delta \lambda^* &= \frac{\lambda _\text{R}^2}{\gamma_1^2}(U - \frac{\Delta}{2} + \frac{\lambda}{2}).
\end{align}
Here, the sign in front of the $U/2$ term matches the value of $\zeta = \pm 1$ introduced in eq \ref{eqn1.2}. In eq \ref{Eqn.peq0b} the plus and minus sign corresponds to the $s = |\uparrow \rangle$ and $s = |\downarrow \rangle$ states respectively. This result extends eq \ref{Eqn.1.5} by including the effects of sublattice asymmetry $\Delta$ and the leading correction at order $\frac{1}{\gamma _1^2}$ given by $\delta \lambda^*$. We see that the effect of $\Delta$ is to uniformly shift both $E_{-1,s}$ bands and renormalize the bias $U$. The quantity $\delta \lambda^*$ produces a similar effect, and can also generate a spin splitting between the $E_{-1,s}$ bands. However, so long as the values $\lambda$ and $\lambda_R$ are comparable, the quantity $\delta \lambda^*$, which is suppressed by a large factor $\frac{1}{\gamma _1^2}$ compared to $\lambda$, only matters as a constant energy shift to the $E_{-1,s}$ bands but not a source of spin splitting.

\begin{figure}
\includegraphics[scale=0.6]{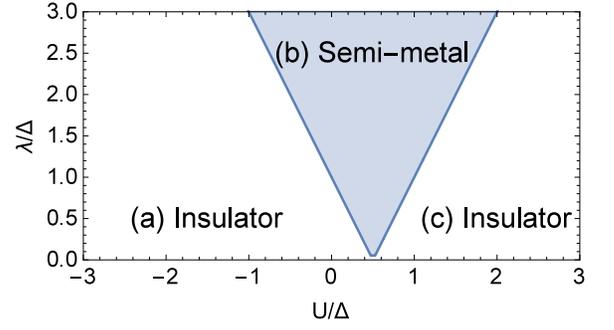}
\caption{\label{phase}Phase diagram of the BLG-on-TMD system described by eq \ref{Eqn.2.2} at charge neutrality when $\lambda \gtrsim \lambda_\text{R}$. Phases (a)-(c) have band structures with corresponding labels in Fig.\ref{figband}.}
\end{figure}

In this case, upon tuning interlayer bias $U$, the $E_{+1,s}$ bands (red) shift across the $E_{-1,s}$ bands (blue) so that the system undergoes transitions from an insulator to a semi-metal and then again to an insulator state. The corresponding phase diagram is shown in Fig.\ref{phase}. At large $U$ (Fig.\ref{figband}(a, c)), the system is insulating at charge neutrality and allows for gate-switching of Ising SOI as discussed above. The sign of $U$ determines which charge carriers, electrons or holes, experience the effective Ising SOI. At not-too-large $U$ values such that $\Delta -\lambda \lesssim 2U \lesssim \Delta + \lambda$, the system is semi-metallic at charge neutrality. As shown in Fig.\ref{figband}(b), in this case the $E_{+1,s}$ bands lie between the $E_{-1,s}$ bands so that the Ising SOI gap is partitioned between the electrons and holes: $\lambda \simeq \Delta E_e + \Delta E_h$. This partitioning can be tuned from $0\%$ to $100\%$ by varying $U$, and thus in this regime both the electron and hole spin splittings are gate-tunable, albeit in a correlated fashion.

As $\lambda$ decreases to zero, its effects at $p=0$ eventually become subleading compared to that of $\delta \lambda^*$ when $\lambda_\text{R} \gg \gamma_1 \sqrt{\frac{\lambda}{U}}$. At $\lambda = 0$, we find from eq \ref{Eqn.peq0b} that $\left. E_{-1,s}\right|_{p=0} \simeq - \frac{U}{2} + \frac{\Delta}{2}+ \delta \lambda^*(1\mp\tau _z)$. The term $\delta \lambda^*(1\mp\tau _z)$ introduces a spin splitting which varies linearly with $U$. In practice, which contribution dominates depends on the actual values of $\lambda$ and $\lambda_\text{R}$; so far, experiments in BLG-on-WS$_2$ indicate that $\delta \lambda^*$ is indeed dominant with $\lambda_\text{R}$ $\sim$ 10 meV\cite{BLGTMD}.

The $\frac{1}{\gamma_1^2}$ suppression in $\delta \lambda^*$ is a consequence of the specific form of $\delta H_\text{R} \propto \tau _z \sigma _x s_y - \sigma _y s_x $. It couples the A1/B2 polarized low-energy states to the high-energy states which are strongly A2-B1 mixed by the interlayer coupling $\gamma_1$. At $p=0$, the wavefunctions of the low-energy states remain layer-polarized such that the SOI is completely tunable. However, because $\delta H_\text{R}$ introduces a substantial interlayer mixing in the low-energy subspace that increases with $p$, the which-layer tunability of the SOI becomes increasingly suppressed away from the valleys as is evident in Fig.\ref{schematic}.

An even more interesting behavior is found when BLG is encapsulated between two TMD layers. In this case, carriers in each of the two graphene layers experience an interfacially-induced SOI from the TMD layers above and below, respectively. The low-energy effective Hamiltonian near the $K_{\pm}$ points now reads
\begin{equation}\label{Eqn.HBLGeff}
H_{\text{eff}} = H_{\text{BLG}} \otimes \mathtt{1}^{(s)} + \sum_{i=1,2}\mathcal{P}_i \delta H_{\text{MLG},i}\mathcal{P}_i,
\end{equation}
in which a layer index $i$ is introduced to allow for distinct phenomenological parameters for the different layers: $\Delta _i, \lambda_i, \lambda_{\text{R},i}$.

A consequence of adding $\delta H_{\text{MLG},2}$ is to open up a gap at $p=0$ between the $E_{+1,s}$ bands. It follows that for arbitrary values of $\Delta _i, \lambda_i, \lambda_{\text{R},i}$, the system at charge neutrality has up to five different insulating phases with phase transitions occurring at values of $U$ for which the band gap at $p=0$ closes. While the overall Chern number for any of these bands is guaranteed to vanish because of time reversal symmetry, the \textit{valley} Chern number is unconstrained and can take non-zero values. In fact, the valley Chern numbers for some of the bands changes across a phase transition, such that these insulating phases are topologically distinct. This suggests that the intrinsic valley Hall conductivity at charge neutrality, $\sigma ^{\text{VH}}_{xy}(0)$, is a suitable quantity to distinguish these phases.

To compute $\sigma ^{\text{VH}}_{xy}(0)$, we first obtain the Berry curvature of each energy band near either valley, $\Omega^{(n)}_\pm (\vec{p})$. Here we introduced a generalized index $n$ that labels bands, including both the four low-energy bands (previously labeled by $\lbrace \zeta, s \rbrace$) as well as the other four high-energy bands that we have excluded from our discussion thus far. Because the Berry curvature is strongly peaked at the valleys, the corresponding valley Chern number $N^{(n)}_\pm$ can be obtained by numerically integrating $\Omega^{(n)}_\pm (\vec{p})$ over an individual valley\cite{numchern}. The intrinsic valley Hall conductivity at charge neutrality is then obtained from adding up the contributions from all the filled bands, $\sigma ^{\text{VH}}_{xy}(0) = \sum_n \left(N^{(n)}_+ - N^{(n)}_-\right) \frac{e^2}{h} = 2\sum_n N^{(n)}_+ \frac{e^2}{h}$. Here we used the relation $N^{(n)}_+ = -N^{(n)}_-$, valid because of time reversal symmetry.

A detailed characterization of the various phases for arbitrary values of $\Delta _i, \lambda_i, \lambda_{\text{R},i}$ lies outside the scope of this work. Here we highlight 
a generic aspect which can be illustrated by considering the simplest case of $\Delta_1 =\Delta_2 = 0$ (this choice of values is consistent with ab initio studies\cite{MLGWS2}), $|\lambda_i| = \lambda$, and $|\lambda_{\text{R},i}| = \lambda_\text{R}$. In this case, the system hosts two topologically distinct phases at charge neutrality -- the ordinary valley Hall phase, in which $\sigma ^{\text{VH}}_{xy}(0) = -4\text{sgn}(U) \frac{e^2}{h}$, and the anomalous valley Hall phase, in which $\sigma ^{\text{VH}}_{xy}(0) \neq -4\text{sgn}(U)\frac{e^2}{h}$. We will denote these phases as VH0 and VH1 respectively. 

A simple way to understand the VH0-VH1 phase transition is as follows. The well-studied case of dual-gated BLG in the absence of SOI, which has $\sigma ^{\text{VH}}_{xy}(0) = -4\text{sgn}(U) \frac{e^2}{h}$\cite{BLGvalley,gappedBLG,valleyexpt,valleyexpt2,valleyexpt3,valleyexpt4}, is in fact a specific example of the VH0 phase for which $\lambda = \lambda_R = 0$. The system remains in the same topological phase VH0 in the presence of relatively weak SOI, i.e. when $\lambda \lesssim |U|$, since the SOI-induced splitting of the low-energy bands is insufficient to cause the band gap to close at either valley. In agreement with the above picture, independent of the relative signs of $\lambda_i$ and $\lambda_{\text{R},i}$, we find that $\sigma ^{\text{VH0}}_{xy}(0) = -4\text{sgn}(U) \frac{e^2}{h}$.

Upon tuning down the interlayer-bias $U$ such that $|U| \lesssim \lambda$, the SOI starts to dominate and band inversion occurs between the low-energy bands at both valleys. The system undergoes a topological phase transition into the VH1 phase through the closing and re-opening of the band gap analogous to phase transitions in Chern and topological insulators. We therefore expect the valley Chern numbers of the low-energy bands to change across the VH0-VH1 phase transition such that $\sigma ^{\text{VH0}}_{xy}(0) \neq \sigma ^{\text{VH1}}_{xy}(0) = 2M\frac{e^2}{h}$, where possible values of $M$ equal $0, \pm 1, \pm 3, ...$; the results from our preliminary studies are consistent with this expectation. Unlike $\sigma ^{\text{VH0}}_{xy}(0)$, the value of $\sigma ^{\text{VH1}}_{xy}(0)$ depends on the relative signs of $\lambda_i$ and $\lambda_{\text{R},i}$. This dependence is, however, not yet well understood. This interplay between the induced SOI and the interlayer bias resembles the behavior for the quantum spin Hall effect in graphene\cite{QSHI1} and the electrically tunable topological insulator\cite{gatedTI}.

In fact, this sharp change in $\sigma ^{\text{VH}}_{xy}(0)$ is independent of the specifics of the induced SOI and occurs in the generic case when $|\lambda_1| \neq |\lambda_2|$ and $|\lambda_{\text{R},1}| \neq |\lambda_{\text{R},2}|$ as well despite there potentially being a more elaborate phase characterization scheme. This change in $\sigma ^{\text{VH}}_{xy}(0)$ no longer happens at $|U| \simeq \lambda$ but at a different critical value $U_\text{c} \lesssim \frac{1}{2}(|\lambda_1| + |\lambda_2|)$. As the experimental study of valley-based transport is still in its infancy, these predictions therefore serve as robust experimental signatures that could be used to advance our understanding of the valley Hall effect. We find ourselves with a system whose topological nature is not completely determined by the material itself, but can instead be gate-controlled, in situ.

In summary, graphene-based heterostructures with on-demand SOI grant access to tunable topological properties. In particular, gate-controlled intrinsic valley Hall conductivity can be achieved in these systems through combining interlayer coupling, gating and various types of interfacially-induced SOI. Further, the robust broad-band response of graphene\cite{dynamic1} turns graphene-on-TMD heterostructures into a unique platform to realize and explore novel time-dependent spin phenomena such as the electrically driven spin resonance. It can also help to extend the optoelectronics and valleytronics  phenomena of current interest\cite{SV2,SV3,SV4,SV5,SV6} into the time domain.

At the time of submitting our manuscript for publication we became aware of Ref.~\citenum{Fab1} that analyzes phenomena closely related to some of those described above. The approach in Ref.~\citenum{Fab1} relies on ab-initio techniques, whereas we develop an approach relying on the low-energy effective Hamiltonian of the system, which provides a direct physical insight into the complexity and richness of our system and pinpoints the key ingredients of the BLG-on-TMD system required for accessing SOI with enhanced gate-tunability.

We thank Zhe Wang and DongKeun Ki for useful discussions. J. K. is supported by the National Science Scholarship from the Agency for Science, Technology and Research (A*STAR). This work was supported by the STC Center for Integrated Quantum Materials, NSF Grant No. DMR-1231319, the SNF, the NCCR QSIT and the EU Graphene Flagship.

\newpage
\section{Appendix: Derivation of low-energy effective Hamiltonian and band structure.}
Here we present the details of our derivation of the low-energy effective Hamiltonian of the BLG-on-TMD system. We first consider the tight-binding Hamiltonian of the (spin-degenerate) bernal-stacked BLG near the $K_\pm$ points. This is given by \cite{epptyBLG}
\begin{equation}
H_{BLG} = 
 \left( \begin{array}{cccc}
\epsilon _{A1} & v\pi ^{\dagger} & -v_4 \pi ^{\dagger} & v_3 \pi \\
v\pi & \epsilon _{B1} & \gamma _1 & -v_4 \pi ^{\dagger} \\
-v_4 \pi & \gamma _1 & \epsilon _{A2} & v\pi ^{\dagger} \\
v_3 \pi ^{\dagger} & -v_4 \pi & v\pi & \epsilon _{B2}
\end{array} \right),
\end{equation}
where $\gamma _1$ is the interlayer (A2-B1) hopping, $v$ is the MLG band velocity, $v_3$ is the velocity associated with trigonal warping and $v_4$ the velocity associated with skew interlayer coupling. $\epsilon _{\alpha i}$ is the on-site energy for sublattice $\alpha = A,B$ and on layer $i = 1,2$. Note that $H_{BLG}$ is expressed in the $(A1 ,B1, A2, B2)$ basis. For simplicity, we neglect the terms associated with $v_3$ and $v_4$ and consider the case with no intrinsic sublattice asymmetry so that $\epsilon _{\alpha i} = \epsilon _i$ with $U=\epsilon _2 - \epsilon _1$. We use this simplified form of $H_{BLG}$ in $H_{\text{eff}}$ defined by eq (6) in the main text.

Since we are only interested in the 4 low-energy bands which have energy scales much smaller than $\gamma _1$, i.e. $|\gamma _1 | \gg |E|, |\lambda |, |\lambda _R |, |U|, |\Delta |$, we can project $H_{\text{eff}}$ to the low-energy subspace and obtain a low-energy effective Hamiltonian for our system. To proceed, we first rewrite it as $H_{\text{eff}} = H_{\gamma _1} + H'$, with
\begin{equation}
H_{\gamma _1} = \left( \begin{array}{cccc}
0 & 0 & 0 & 0 \\
0 & 0 & \gamma _1 & 0 \\
0 & \gamma _1 & 0 & 0 \\
0 & 0 & 0 & 0
\end{array} \right) \otimes \mathtt{1}_2 ^{(s)},
\end{equation} 
and re-express it in the basis that diagonalizes $H_{\gamma _1}$:
\begin{equation}
\tilde{H}_{\text{eff}} = \tilde{H}_{\gamma _1} + \tilde{H'}, \quad \tilde{H}_{\gamma _1} = \left( \begin{array}{cccc}
0 & 0 & 0 & 0 \\
0 & 0 & 0 & 0 \\
0 & 0 & -\gamma _1 & 0 \\
0 & 0 & 0 & \gamma _1
\end{array} \right) \otimes \mathtt{1}_2 ^{(s)}.
\end{equation} 

In this basis, we can label each eigenstate by a quantum number $m = 0, \pm1$ such that $\tilde{H}_{\gamma _1} |m \rangle = m \gamma _1 |m \rangle $. The low-energy subspace that we are interested in is spanned by the $m = 0$ eigenstates. We can then decompose $\tilde{H'}$ as
\begin{equation}
 \tilde{H'} = T_0 + T_1 + T_{-1} + T_2 + T_{-2},
\end{equation}
where the operators (Unitary but not necessarily Hermitian) $T_i = T_{-i}^\dagger$ can be understood as ladder operators -- $T_i |m \rangle \varpropto |m + i\rangle $ -- and are given by
\begin{eqnarray}
T_0 &\equiv & \frac{1}{2} \left( \begin{array}{cccc}
-U \mathtt{1}_2 ^{(s)} + 2 w_+ & 0 & 0 & 0 \\
0 & U \mathtt{1}_2 ^{(s)} & 0 & 0 \\ 
0 & 0 & w_- & 0 \\
0 & 0 & 0 & w_-
\end{array} \right), \nonumber \\
T_1 &\equiv & \frac{1}{\sqrt{2}} \left( \begin{array}{cccc}
0 & 0 & v\pi ^\dagger \mathtt{1}_2 ^{(s)} + a_R^\dagger & 0 \\
0 & 0 & -v\pi \mathtt{1}_2 ^{(s)} & 0 \\
0 & 0 & 0 & 0 \\
v\pi \mathtt{1}_2 ^{(s)} + a_R & -v\pi ^\dagger \mathtt{1}_2 ^{(s)} & 0 & 0
\end{array} \right), \nonumber \\
T_2 &\equiv & \frac{1}{2}\left( \begin{array}{cccc}
0 & 0 & 0 & 0 \\
0 & 0 & 0 & 0 \\
0 & 0 & 0 & 0 \\
0 & 0 & -U \mathtt{1}_2 ^{(s)} +w_- & 0
\end{array} \right),
\end{eqnarray}
where $w_\pm=\frac{1}{2}(\pm\Delta \mathtt{1}_2 ^{(s)} + \lambda \tau _z s_z)$, $a_R=\frac{1}{2} \lambda _R (\tau _z s_y - i s_x)$.

With this framework in place, we proceed with the Brillouin-Wigner method outlined in Section IV of Ref.~\citenum{Heff} to obtain the effective Hamiltonian to arbitrary orders in $\frac{\Lambda}{\gamma _1} $, where $\Lambda $ refers to any energy scale set by the other parameters (including $v p$). In this expansion scheme, at second order, we recover the usual spin degenerate effective 2-band BLG Hamiltonian $H^{2 \times 2}_{\text{BLG}}$. The correction due to SOI at the same order ($H^{4 \times 4}_{\text{SO, 1}}$) is off-diagonal, the energy corrections of which only enter at the next order. To consistently account for the energy corrections to the $H^{2 \times 2}_{\text{BLG}}$ spectrum, we therefore go to third order in the expansion scheme and obtain
\begin{eqnarray}\label{Eqn.2.6}
H_{\text{eff}}^{4 \times 4} &\approx & H^{2 \times 2}_{\text{BLG}} \otimes \mathtt{1}_2 ^{(s)} + H^{4 \times 4}_{\text{SO, 0}} + H^{4 \times 4}_{\text{SO, 1}} + H^{4 \times 4}_{\text{SO, 2}} \nonumber \\
&+& \gamma _1 O\left(\frac{\Lambda}{\gamma _1} \right)^4, \quad\text{with} \\
H^{2 \times 2}_{\text{BLG}} &\equiv &\left( \begin{array}{cc}
-\frac{1}{2} U & -\frac{1}{\gamma _1} \left( v \pi ^\dagger \right)^2 \\
-\frac{1}{\gamma _1} \left( v \pi \right)^2 & \frac{1}{2} U
\end{array} \right), \nonumber \\
H^{4 \times 4}_{\text{SO, 0}} &\equiv & \left( \begin{array}{cc}
w_+ & 0 \\
0 & 0
\end{array} \right), \nonumber \\
H^{4 \times 4}_{\text{SO, 1}} &\equiv & -\frac{1}{\gamma _1} \left( \begin{array}{cc}
0 & v \pi ^\dagger a_R^\dagger \\
v \pi a_R & 0
\end{array} \right), \nonumber \\
H^{4 \times 4}_{\text{SO, 2}} &\equiv & \frac{1}{2 \gamma _1^2}\left( \begin{array}{cc}
A & 0 \\
0 & B
\end{array} \right), \nonumber \\
A &\equiv & \frac{1}{2}\lambda _R^2 (\mathtt{1}_2 ^{(s)}-\tau _z s_z)(2U - \Delta + \lambda) \nonumber \\
&+&(2U - \Delta) v \lambda _R (p_x s_y - p_y s_x) + 2(U \mathtt{1}_2 ^{(s)} - w_+ ) v^2 p^2, \nonumber \\
B &\equiv & 2(-U \mathtt{1}_2 ^{(s)} + w_-) v^2 p^2. \nonumber
\end{eqnarray}

Here, we have explicitly separated the resulting effective Hamiltonian to the usual spin degenerate effective 2-band BLG Hamiltonian $H^{2 \times 2}_{\text{BLG}}$ and the corrections due to substrate induced SOI $H^{4 \times 4}_{\text{SO, j=0,1,2}}$. These corrections are further organized into powers of $\gamma _1^{-1}$, so that the $(j-1)^{th}$ order correction $H^{4 \times 4}_{\text{SO, j-1}} \varpropto \gamma _1 ^{-j+1}$.

Solving for the eigenvalues of $H^{4 \times 4}_{\text{eff}}$ perturbatively, we obtain the low-energy band structure to order $O\left(\frac{\Lambda ^3}{\gamma _1^2} \right)$ for $\tau_z = 1$: 
\begin{widetext}
\begin{eqnarray}
E_{+1,|\uparrow \rangle} &=& \frac{U}{2} + \frac{v^2 p^2}{\gamma _1 ^2} \left( - U - \frac{\Delta}{2} + \frac{\lambda}{2} + \frac{2\lambda _R ^2}{2U - \Delta + \lambda } + \frac{2v^2 p^2}{2U - \Delta - \lambda } \right) + O\left(\frac{\Lambda ^4}{\gamma _1^3} \right), \nonumber \\
E_{+1,|\downarrow \rangle} &=& \frac{U}{2} + \frac{v^2 p^2}{\gamma _1 ^2} \left( - U - \frac{\Delta}{2} - \frac{\lambda}{2} + \frac{2\lambda _R ^2}{2U - \Delta - \lambda } + \frac{2v^2 p^2}{2U - \Delta + \lambda } \right) + O\left(\frac{\Lambda ^4}{\gamma _1^3} \right), \nonumber \\
E_{-1,|\uparrow \rangle} &=& - \frac{U}{2} + \frac{\Delta + \lambda}{2} + \frac{v^2 p^2}{\gamma _1 ^2} \left(U - \frac{\Delta}{2} - \frac{\lambda}{2} - \frac{2\lambda _R ^2 + 2v^2 p^2}{2U - \Delta - \lambda } \right) + O\left(\frac{\Lambda ^4}{\gamma _1^3} \right), \nonumber \\
E_{-1,|\downarrow \rangle} &=& - \frac{U}{2} + \frac{\Delta - \lambda}{2} + \frac{1}{\gamma _1 ^2} \left\lbrace \lambda _R ^2 \left(U - \frac{\Delta}{2} + \frac{\lambda}{2} \right) \right. \nonumber \\
&& + \left. v^2 p^2 \left(U - \frac{\Delta}{2} + \frac{\lambda}{2} - \frac{2\lambda _R ^2}{2U - \Delta + \lambda } \right) -  v^4 p^4 \left( \frac{2}{2U - \Delta + \lambda } \right) \right\rbrace + O\left(\frac{\Lambda ^4}{\gamma _1^3} \right).\nonumber \\ 
\end{eqnarray}
\end{widetext}

\input{LayertuneSO_arXiv_submit.bbl}
%\bibliography{bibarXiv}

\end{document}